\input{aipcheck}
\documentclass[
    ,final             
  ]
  {aipproc}
\layoutstyle{6x9}
\usepackage{epsfig}
\begin{document}
\def\be{\begin{equation}}
\def\ee{\end{equation}}
\def\ba{\begin{eqnarray}}
\def\ea{\end{eqnarray}}
\def\Mesz{M\'esz\'aros}
\def\siml{\lower4pt \hbox{$\buildrel < \over \sim$}}
\def\simg{\lower4pt \hbox{$\buildrel > \over \sim$}}
\def\etal{{\it et al.}}
\def\msun{M_\odot}
\def\eps{\epsilon}
\newcommand{\figuresize}{0.41\textwidth}
\newcommand{\boxsize}{0.89\textwidth}
\newcommand{\smallboxsize}{0.8\textwidth}
\title{Electromagnetic Signals from Planetary Collisions}

\author{Bing Zhang and Steinn Sigurdsson }
{
  address={
Department of Astronomy \& Astrophysics,
Penn State University, University Park, PA 16802}
}
\begin{abstract}
We investigate the electromagnetic signals accompanied with 
planetary collisions and their event rate, and explore the possibility
of directly detecting such events. A typical Earth--Jupiter collision
would give rise to a prompt EUV-soft-X-ray flash lasting for hours and
a bright IR afterglow lasting for thousands of years. 
With the current and forthcoming observational technology and
facilities, some of these collisional flashes or the post-collision
remnants could be discovered. 
\end{abstract}
\maketitle

\section{Introduction}

More than 100 extra-solar planets have been detected. 
At the same time, our understanding of astrophysical 
phenomena has been greatly boosted through studying cataclysmic
transient events such as supernovae, X-ray bursts, and gamma-ray
bursts. Another type of electromagnetic transient events that arises
from collisions of extra-solar planets has been long predicted in the
context of planet formation\cite{wetherill90}. Recently we
discussed the possible electromagnetic signals accompanying such
collisional events\cite{zs03}. Discovering such events would
undoubtedly have profound implications of understanding the formation
rate, dynamical instability, as well as internal composition and
structure of extrasolar planets.

\section{Electromagnetic Signals: Three Stages}

Generally one can categorize the collision-induced signals into three
stages, defined by three characteristic time scales. Below we will
take a head-on collision between a Jovian planet and an
Earth-size planet as an example. 
A schematic lightcurve is presented in Figure 1.

 \begin{figure}[htb]
 \begin{minipage}[t]{0.9 \textwidth}
 \epsfxsize=\boxsize
 \epsfbox{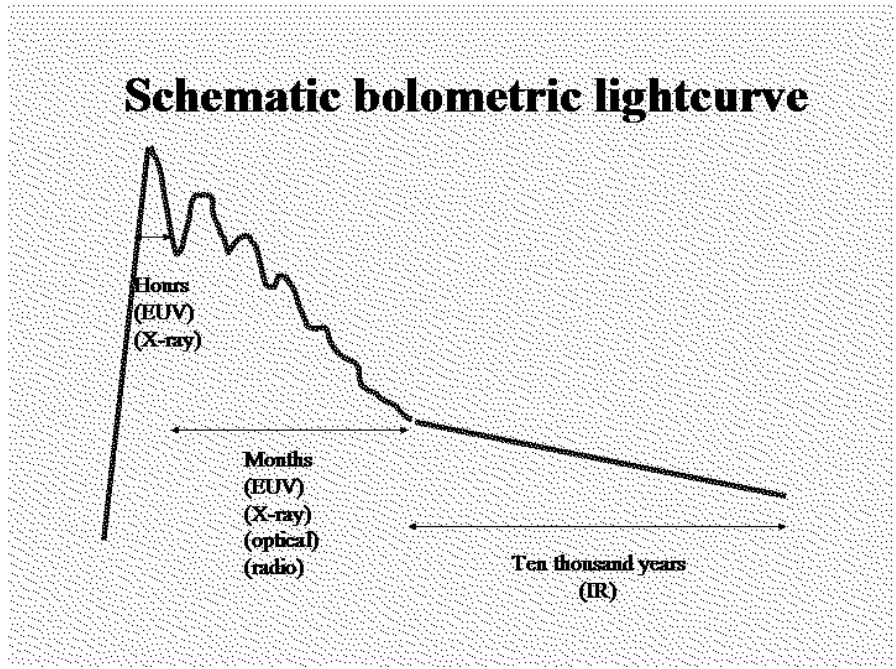}
 \end{minipage}
 \caption{A schematic bolometric lightcurve for a Jupiter-Earth
 collision event. Three stages are highlighted: (1) a prompt
 EUV-soft-X-ray flash, characterized by a sharp rise and milder decay
 lasting for hours; (2) a spin-modulated decaying lightcurve lasting 
 for about a month, which could be visible in the EUV, X-ray, optical
 and radio bands; (3) a long-term IR warm afterglow lasting for
 thousands of years.} 
 \label{fig:lightcurve}
 \end{figure}

\subsection{Stage 1: Prompt EUV-Soft-X-ray Flash}

Assuming zero velocity at infinity, the total energy of the collision
is about $6\times 10^{40} ~ {\rm erg}$. Upon impact, a reverse shock
propagates into the impactor (i.e. the Earth-size planet), and the
shock crossing time scale is $\sim 10$ minutes.
This defines the rising timescale of the collision-induced
electromagnetic flash. The energy deposition rate during this stage is
$\sim 10^{38}~{\rm erg~s^{-1}}$, much greater than Jupiter's Eddington 
luminosity. A likely picture is that the prompt heat generated upon
collision would dissociate the molecules and ionize the atoms within a 
short period of time, and the emission quickly becomes
Eddington-limited. After the peak, the lightcurve decays mildly as the
deceleration of the impactor is still going on inside the Jovian
planet. The timescale of the prompt flash can be estimated to be of
order of 10 times of the rising time, i.e.,
\be 
\tau_1 \sim 2~{\rm hr}.
\ee
The bolometric luminosity is near Eddington, i.e. $L_{pk} \sim 5\times
10^{34}~{\rm erg~s^{-1}}$,
with a thermal temperature $T_{pk} \sim 1.1\times 10^5 ~{\rm K}$
emitting from a hot spot with a radius comparable of the Earth radius.
The peak flux is $F_\nu (pk) \sim 60~{\rm \mu Jy}~ (D/{\rm 
10 kpc})^{-2}$.
peaking in the EUV band. A non-thermal tail due to Comptonization would
extend into the soft-X-ray band, detectable through out the Galaxy if
the neutral hydrogen absorption is not important, and is even visible
from some nearby galaxies. The prompt flash greatly increases the
planet-to-star flux ratio, making them detectable in the optical band
through photometry monitoring [$f({\rm U, pk})\sim 0.2$, $f({\rm
V,pk}) \sim 0.02$, and $f({\rm I,pk}) \sim 0.008$, where $f(\nu)
\equiv {F(\nu,{\rm planet})}/{F(\nu,{\rm star})}$].
The total energy radiated during this prompt phase is a tiny fraction 
($\sim 0.5\%$) of the total energy deposited. The majority of energy
is stored as latent heat and radiated over a much longer time scale.

\subsection{Stage 2: Spin-Modulated Decaying Phase}

After the prompt phase ends (i.e. the impactor is stopped inside the
giant planet), the luminosity steadily drops. The heat deposited deep
inside the giant planet would excite a vigorous convective flow. The
area of the hot spot gradually gets larger and larger until the whole
surface reaches the same temperature. The time scale for retaining a hot 
spot could be estimated as
\be
\tau_2 \sim 1~{\rm month}.
\ee
During this time, an observer would see some quasi-periodic signal due 
to the modulation of the planet spin (with a period $\siml 1$ day), as 
the hot spot enters and leaves the field of view. The modulation
pattern could be visible in EUV-X-ray, or in optical through
photometric monitoring, or in radio. A radio flare is expected during
the vigorous convective epoch due to the enhanced dynamo activity
inside the giant planet. The periodic pattern would be gradually
smeared out as the hot spot boundary gradually increases.

\subsection{Stage 3: Long-Term IR Afterglow} 

After the surface temperature becomes uniform, the giant planet keeps
cooling, radiating away the majority of the energy deposited during
the impact. This time scale is much longer, typically
\be
\tau_3 \sim 10^3-10^4 ~ {\rm yr}.
\ee
The channel of emission is mainly in IR\cite{stern94}. During this
epoch, the planet-to-star flux ratio in the IR band is very high.
For a G2 host star, the typical I- and
K-band flux contrasts are $f({\rm I,ag}) \sim 2.6\times
10^{-4}$ and $f({\rm K,ag}) \sim 1.7\times 10^{-3}$. This is favorable 
to be detected in nearby star forming regions\cite{stern94}.

\section{Detectability} 

Numerical simulations\cite{cwb96,fhr01} suggest that collisions of the
type we are discussing are plausible. The basic picture is that there
are secular perturbations of the inner planets, which over time scales
comparable to the age of the system lead to large
changes in eccentricity and semi-major axis for one or more planets,
leading to a large probability of collision. 

The event rate can be roughly estimated. Assuming that on average a
solar-system like our own has 5 collisions over its life time, and
that essentially every star harbors a planetary system, the collision
event rate in our galaxy would be about $5\times 10^{11}/(10^{10}~{\rm 
yr}) \sim 50/{\rm yr}$. 

Considering an ensemble of stars with an average age $\bar t$, in
order to detect one 
event with duration $\tau$ after a continuous observation time of
$t_{obs}$, the critical number of stars in this ensemble that have to
be searched is 
\be
N_{*} = (f_p \bar N_c)^{-1} \frac{\bar t}{{\rm max}(\tau,t_{obs})}, 
\ee
where $f_p \sim 100\%$ is the fraction of the stars in the ensemble
that have planets, and $\bar N_c \sim 5$ is the average total number
of collisions during the lifetime of a typical star in the ensemble. 
One can also define a characteristic flux $F_{\nu,c}$
of the collisional events. Given a number density $n_*$ of the stars
with the average age $\bar t$, one can estimate a critical distance
one has to search in order to find one collisional event, i.e., $D_c
\sim (N_*/n_*)^{1/3}$. The typical flux can be then estimated with
$D_c$. Although $N_*$ is sensitive to the average age $\bar t$ of the
ensemble of stars being investigated, $D_c$ and $F_{\nu,c}$ are
essentially independent of the ensemble adopted, given a constant
birth rate of stars. According to such an estimate, the {\em Extreme
Ultraviolet Explorer (EUVE)} all-sky survey\cite{mcdonald94} may have
recorded $\sim 10$ such events, with the brightest one having a count 
rate of about $(170-680)~{\rm counts~ks^{-1}}$ at 100 angstrom. 
The predicted flux level is also well above the
sensitivity of soft X-ray detectors, such as {\em ROSAT}, {\em
Chandra}, and {\em XMM-Newton}, so that there might be such events
recorded in their archival data as well.

\section{Searching strategy}

A future dedicated wide field detector sensitive to (50-200) angstrom
would be able to detect 10's of EUV-soft-X-ray flares per year due to
planetary collisions. 
Photometrically monitoring a huge number of stars over a long period
of time by missions such as {\em GAIA}\cite{perryman01} (or other
synoptic all-sky surveys) for several years should lead to detections
of several collisional events. 
If an EUV-soft-X-ray flare is detected, a search of its optical and
radio counterpart (like catching afterglows in gamma-ray 
burst study) is desirable. A planetary-rotation-period
modulated fading signal would be an important clue. Doppler radial
velocity measurements and IR monitoring may be
performed later for the collision candidate to verify the existence of
the planet(s). Each collisional event would leave a bright remnant
glowing in IR. The duration of the afterglow is thousands of 
years. Such afterglows could be directly searched in the nearby star
forming regions, and the planets could be directly imaged.

\begin{theacknowledgments}
This work is supported by NASA NAG5-13286, the NSF grant
PHY-0203046, the Center for Gravitational Wave Physics (under NSF PHY
01-14375), and the Penn State Astrobiology Research Center.
\end{theacknowledgments}


\bibliographystyle{aipproc}   

\begin{thebibliography}{99}

\bibitem{wetherill90} Wetherill, G. W. Ann. Rev. Earth Planet. Sci.,
{\bf 18}, 205 (1990)

\bibitem{zs03} Zhang, B. \& Sigurdsson, S. S. ApJ, {\bf 596}, L95
(2003) 

\bibitem{stern94} Stern, S. A. AJ, {\bf 108}, 2312 (1994)

\bibitem{cwb96} Chambers, J. E., Wetherill, G. W., \& Boss, A. P. 
Icarus, {\bf 119}, 261 (1996)

\bibitem{fhr01} Ford, E. B., Havlickova, M. \& Rasio, F. A. Icarus, 
{\bf 150}, 303 (2001)

\bibitem{mcdonald94} McDonald, K. et al. AJ, {\bf 108}, 1843 (1994)

\bibitem{perryman01} Perryman, M. A. C., et al., A\&A, {\bf 369}, 339
(2001) 

\end{thebibliography}



\IfFileExists{\jobname.bbl}{}
 {\typeout{} \typeout{******************************************}
  \typeout{** Please run "bibtex \jobname" to optain} \typeout{** the
  bibliography and then re-run LaTeX} \typeout{** twice to fix the
  references!}  \typeout{******************************************}
  \typeout{} }

\end{document}